# High-resolution absorption measurements with free-electron lasers using ghost spectroscopy

Yishai Klein[1], Edward Strizhevsky[1], Flavio Capotondi [2], Dario De Angelis [2], Luca Giannessi[2,7], Matteo Pancaldi[2], Emanuele Pedersoli[2], Giuseppe Penco[2], Kevin C. Prince[2,3], Or Sefi[1], Young Yong Kim[4,5], Ivan A. Vartanyants[4,6] and Sharon Shwartz[1]

[1] *Physics Department and Institute of Nanotechnology and advanced Materials, Bar Ilan University, Ramat Gan, 52900, Israel*
[2] *Elettra-Sincrotrone Trieste, Strada Statale 14-km 163.5, Basovizza, 34149, Trieste, Italy*
[3] *Department of Chemistry and Biotechnology, School of Science, Computing and Engineering Technology, Swinburne University of Technology, Melbourne, VIC, Australia*
[4] *Deutsches Elektronen-Synchrotron DESY, Notkestr. 85, 22607 Hamburg, Germany*
[5] *European XFEL, Holzkoppel 4, 22869 Schenefeld, Germany*
[6] *National Research Nuclear University MEPhI (Moscow Engineering Physics Institute), Kashirskoe shosse 31, 115409 Moscow, Russia*
[7] *Istituto Nazionale di Fisica Nucleare, Laboratori Nazionali di Frascati, Via E. Fermi 54, 00044 Frascati (Roma), Italy*

E-mail: sharon.shwartz@biu.ac.il

**Abstract**

We demonstrate a simple and robust high-resolution ghost spectroscopy approach for x-ray and extreme ultraviolet absorption spectroscopy at free-electron laser sources. Our approach requires an on-line spectrometer before the sample and a downstream bucket detector. We use this method to measure the absorption spectrum of silicon, silicon carbide and silicon nitride membranes in the vicinity of the silicon $L_{2,3}$-edge. We show that ghost spectroscopy allows the high-resolution reconstruction of the sample spectral response using a coarse energy scan with self-amplified spontaneous emission radiation. For the conditions of our experiment the energy resolution of the ghost-spectroscopy reconstruction is higher than the energy resolution reached by scanning the energy range by narrow spectral bandwidth radiation produced by the seeded free-electron laser. When we set the photon energy resolution of the ghost spectroscopy to be equal to the resolution of the measurement with the seeded radiation, the measurement time with the ghost spectroscopy method is shorter than scanning the photon energy with seeded radiation. The exact conditions for which ghost spectroscopy can provide higher resolution at shorter times than measurement with narrow band scans depend on the details of the measurements and on the properties of the samples and should be addressed in future studies.

## 1. Introduction

X-ray and extreme ultraviolet (XUV) free-electron lasers (FELs) are very powerful and bright sources that enable measurements of ultrafast phenomena in a broad range of processes [1,2]. Short wavelength spectroscopy is widely used for the determination of the electronic structure of materials and provides element specific information on the charge and spin structures as well as bonding configurations, which are important for understanding the functionality of materials [3]. When performed at FELs, x-ray spectroscopy can provide information on the dynamics of the processes by using pump-probe schemes, where the short FEL pulse probes a process that is triggered by a pump stimulus, which can be provided either by an optical laser or by the FEL itself. By varying the delay between the pump and the probe pulses, full information on the dynamics of the electronic response of the sample can be recorded [4].

There are basically two common strategies for the measurement of high-resolution absorption spectra. The first is to use narrowband (quasi-monochromatic) radiation and to measure the total transmitted intensity after the sample (or the emitted fluorescence, which is proportional to the absorption). With this approach the spectrum of the sample response is reconstructed by scanning the photon energy of the input radiation and registering the intensities measured by the detector for each input photon energy. The monochromaticity of the pulse is obtained either by using a monochromator [5] or by using one of the seeding schemes depending on the wavelength of the radiation [6–8]. A second strategy is implemented when the radiation has a broadband emission spectrum ($\Delta\lambda/\lambda \sim 1\%$ or more). In this case, the spectrum of the transmitted radiation is compared with the spectrum of the input beam before the sample [9,10] and it is absolutely necessary to know the spectrum before and after the sample with high precision and fidelity. The energy resolution of the first approach depends on the spectral bandwidth of the input radiation whereas, in the second case, it is determined by the resolving power of the spectrometers that are used for the spectral measurements.

The advantages of the broad bandwidth pulse strategy are the possibility to measure broad ranges of spectra without scanning the central emission wavelength and the availability of higher flux. Therefore, this approach can be significantly faster than the narrow bandwidth approach and useful for the measurement of low efficiency processes. However since broad bandwidth FEL pulses are generated usually by using the process of self-amplified spontaneous emission (SASE) [11,12], the pulse energy and the spectra vary randomly from one shot to another. Thus, it is necessary to measure the spectra before and after the sample on a shot-to-shot basis. While single shot spectrometers have been developed [13–17], the simultaneous application of two such spectrometers for the measurement is very challenging and time consuming and the spectrometers are expensive. Furthermore, signal-to-noise (SNR) requirements impose a limitation on the minimum number of photons that must be detected for the reconstruction of the spectra, which leads to stringent requirements for the input flux and limits the dynamical range of absorption magnitudes that can be measured. Thus, the range of samples that can be measured with standard methods is limited. Finally, while with a narrowband pulse the absorption can be inferred from the measurement of the fluorescence yield, which is proportional to the absorbance, in the broadband scheme it is not possible since the comparison between the two spectrometers is required. Thus, the scheme can only be applied to transmissive samples, which strongly limits the choice of materials that can be studied.

An alternative strategy to perform absorption spectroscopy with FEL radiation that overcomes the challenges of the above-described approaches is ghost spectroscopy (GS), which is a form of correlation spectroscopy. This technique has been demonstrated recently with soft x-rays at the Linac Coherent Light Source (LCLS) [18,19]. The concept of GS is closely related to ghost imaging (GI) [20], which has been successfully applied with laboratory [21,22], synchrotron [23–27], and FEL [28] X-ray sources.

The key parameter for GS is the variation of the spectral features from one shot to another. GS indeed exploits the stochastic nature of the SASE pulse spectra, i.e. the random shot-to-shot variation of the multi-spike spectra. Within this method, the spectrum of radiation impinging on the sample is measured and correlated on a shot-by-shot basis with the measured intensity of a single-pixel detector (usually a photodiode) that has no spectral resolution and is mounted after the sample. The measured intensity at this detector is proportional to the integral of the product of the spectrum of the input pulse and the spectral dependence of the transmission of the sample (the transmission function of the sample). Therefore, for each pulse, if the correlation between the input spectrum and the transmission function of the sample is high, the detector measures high intensity. Conversely, if the correlation is low, it measures low intensity. By repeating this procedure for many input pulses with different spectral distributions, it is possible to reconstruct the absorption spectrum of the sample [29,30]. The term ghost here refers to the fact that neither of the detectors can provide the spectrum, directly in analogy to GI, where the bucket detector does not provide spatial information [20–28].

Here, we present a simple and robust approach for GS in the XUV photon-energy range that requires only one spectrometer in front of the sample and a single-pixel detector without any spectral resolution placed behind it. By directly comparing the measurement times and the spectral resolution of the GS case with that obtained by setting the FEL emission in SASE and seeded configurations, we demonstrate that GS is an efficient strategy to perform absorption spectroscopy at FELs.

## 2. Methods

### 2.1 Experimental setup and radiation properties

We conducted the experiment at the DiProl end station [31,32] using the double cascade FEL source FEL-2 of the FERMI user facility located in Trieste, Italy [33]. This source can produce either SASE FEL radiation [34] or seeded FEL pulses [7,33] depending on the setting parameters. To demonstrate our approach for GS, we tuned the SASE pulse central energy in the photon energy range between 99 eV and 106 eV for the measurements of the Si $L_{2,3}$ edges. The radiation produced by the source was focused by a set of bendable Kirkpatrick–Baez mirrors [35] to a spot size of about $500\times600$ μm$^2$ at the sample position, the polarization was circular and the pulse duration was estimated to be about 250 fs. The repetition rate was 50 Hz. The setup of our experiment is presented in figure 1. The on-line spectrometer was the Pulse-Resolved Energy Spectrometer Transparent and Online (PRESTO) [17], which is mounted at FERMI after the undulators and before the end stations. In the PRESTO spectrometer, a grating delivers most of the radiation in zeroth order (97%) to the end-stations, while the weaker first order of the grating is used to measure the spectrum of each pulse. The spectrometer resolution in the working energy range is $\Delta\lambda/\lambda \sim 5 \times 10^{-5}$ [17], corresponding to an energy resolution of about 5 meV at 100 eV. Examples of the spectral distribution of the FEL pulses in the SASE configuration and the average over 8,000 shots are presented in figure 2(a). In our experiment, we mounted the sample in the direct beam and measured three different membranes of silicon (Si), silicon nitride ($Si_3N_4$), and silicon carbide (SiC), 200 nm thick provided by Norcada®. The average energy per pulse at the sample plane was 18.5±3 μJ, corresponding to a deposited energy density per pulse of 6±1 mJ/cm$^2$, which is well below the typical damage threshold of the samples [36]. For the detector with no energy resolution, we used a 10×10 mm$^2$ photodiode with a YAG scintillator screen to convert the FEL radiation into optical radiation. The detector was mounted at about 500 mm downstream from the sample. At the detector position, due to the optical beam divergence, the FEL spot size was about 3×4 mm$^2$.

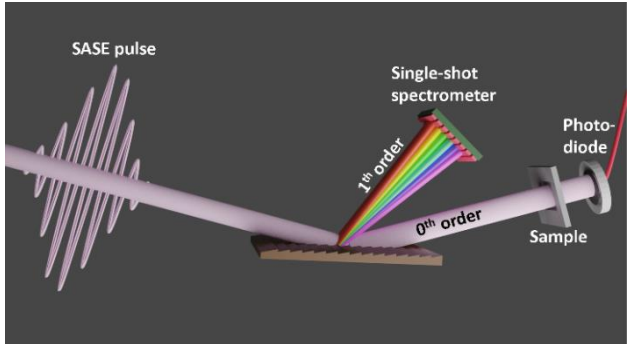

*Figure 1 – GS experimental scheme. The input SASE radiation is split by a grating installed at the PRESTO instrument. The first order is used as a reference that is measured by a single-shot spectrometer mounted before the sample. The zeroth order irradiates the sample and the transmitted radiation after the sample is measured by a photodiode. See further details in the text.*

For the GS measurements and photon energies in the range of 99 eV to 106 eV we varied the central photon energy of the SASE radiation with a step size of 250 meV. At each of the SASE photon energies we measured 2000 shots and the SASE bandwidth spans a range of 500 meV (FWHM, as can be seen in figure 2(a)).

To account for the dependence of the beamline transmission on the photon energy, we compared the total intensities of the spectrometer and the photodiode without the sample every time we changed the central photon energy of SASE emission. We eliminated the background noise of the camera of the spectrometer by subtracting the dark reference images collected without FEL illumination.

Since GS is based on the intensity correlation between the spectral features recorded on the spectrometer and the intensity fluctuation recorded by the single-pixel detector placed behind the sample, it was important to ensure a high linear correlation between the total recorded intensities on the two detectors in the absence of the sample. To compare the correlation quality between the two devices, we define the relative error 'R' for each i$^{th}$ pulse as:

(1) $$R_i = \frac{I_{S_i}/\langle I_S \rangle - I_{P_i}/\langle I_P \rangle}{I_{P_i}/\langle I_P \rangle},$$

where $I_{S_i}$ and $I_{P_i}$ are the total intensity of i$^{th}$ shot, measured by the spectrometer and the photodiode respectively, and $\langle ... \rangle$ represents an average over all pulses.

We found that the typical standard deviation of the relative error distribution of our experimental setup was about 7.5%. A typical distribution of the relative intensity errors between the two detectors, for an ensemble of 36,000 FEL pulses at a variety of energies within our scan range, is presented in figure 2(b).

## 2.2 Seeded radiation and synchrotron measurements

To compare the method of GS with scanning approaches, we used FERMI FEL radiation in the seeded configuration, where the FEL pulses are generated using a ~100 fs (FWHM) external laser pulse in the ultraviolet range from 4.7 to 4.9 eV to trigger the FEL amplification process [7,33]. By using a harmonic upshift factor of 21 the output radiation spans a photon energy from 99 eV to 103 eV. As shown in [37] the seeded FEL generated at FERMI is close to the Fourier limit and the pulse duration is estimated to be about 30 fs [38]. In this experiment the normalized FWHM bandwidth is $\Delta\lambda/\lambda = 1 \times 10^{-3}$, equivalent to an energy resolution of about 100 meV at 100 eV. Therefore, we scanned the photon energy of the seeded radiation with step sizes of 75 meV and measured the transmission of the silicon membrane near the $L_{2,3}$ edges in a range comparable to the GS measurement. The transmission of the membrane at each photon energy was calculated as the ratio between the average pulse energy detected by the photodiode and the average total pulse energy detected by the spectrometer. In addition, we performed the same procedure for the SASE scanning: in this case, almost all the electron bunch participates in the FEL process, resulting in a FEL pulse duration of about 250fs (FWHM).

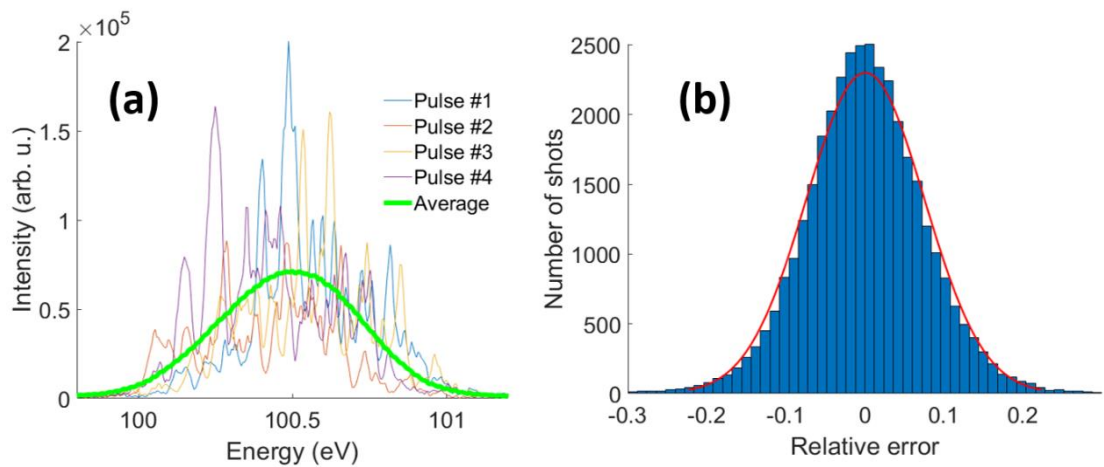

*Figure 2 – (a) Energy distribution of the SASE pulses measured by the on-line spectrometer, when the central photon energy is set at 100.5 eV. The thinner lines are the spectra of 4 individual pulses and the thick green line is the average over 8000 pulses. (b) The relative error distribution between the total intensities at the spectrometer and the photodiode averaged over 36,000 pulses (blue bins). The red line is the Gaussian fit and the standard deviation is 7.5%.*

To validate our method, we compared the spectra we measured at FERMI with the spectral measurements of the same samples at the BEAR beamline at the Elettra synchrotron [39]. Here the transmission at each energy point is simply the average intensity measured after the sample divided by the average intensity before the sample.

### 2.3 GS reconstruction procedure

To reconstruct the ghost spectrum for each SASE central energy, we exploited the following reconstruction procedure. We represent the intensities of the N pulses measured by the photodiode by a vector **T** (test data). The spectra of the pulses are represented by the matrix **A** for which every row is the spectral distribution of a single pulse (reference data). We represent the transmission function of the sample as a vector **x**, and thus the vector **T** is equal to the product of the matrix **A** and the vector **x**

$$(2) \quad \mathbf{Ax} = \mathbf{T}.$$

In GS experiments, we measure the vector **T** and the matrix **A**. We are interested in solving Eq. (2) for the vector **x** using the compressive sensing (CS) algorithm of "total variation minimization by augmented Lagrangian and alternating direction algorithms" (TVAL3) [40]. However, this algorithm works well only when the width of the average spectrum is much broader than the spectral range under investigation. Unfortunately, the SASE bandwidth in our case was narrower than the total measured spectral range. To overcome this challenge, we used the following procedure: first we normalized the matrix of the raw data by the average SASE distributions (for example, for the central photon energy at 100.5 eV we used the green line in figure 2(a))

$$(3) \quad \mathbf{B}_{i,j} = \frac{A_{i,j}}{\frac{1}{N}\sum_{k=1}^{N} A_{k,j}}.$$

Next, we normalized the reference and the test data by the pulse energy of each pulse

$$(4) \quad \mathbf{C}_{i,j} = \frac{B_{i,j}}{\frac{1}{M}\sum_{k=1}^{M} B_{i,k}} \;,\quad \mathbf{T'}_i = \frac{T_i}{\frac{1}{M}\sum_{k=1}^{M} B_{i,k}}.$$

By using this procedure, we can replace Eq. (2) with a new equation

$$(5) \quad \mathbf{Cx'} = \mathbf{T'},$$

where the matrix representing the different energy distributions in each pulse is now the effective matrix **C** where the envelope is normalized and the shot-to-shot intensity variations are filtered out.

Next, we used the TVAL3 algorithm to solve Eq. (5). The basic idea of TVAL3 is to recognize that the gradients of the measured spectra can be represented by a sparse vector. The vector **x'** is reconstructed by minimizing the augmented Lagrangian

$$(6) \quad \min_{\mathbf{x'}} \sum_{j=1}^{M} \|D_j \mathbf{x'}\|_2 + \frac{\mu}{2}\|\mathbf{Cx'} - \mathbf{T'}\|_2^2 \;\; subject\; to \;\; \mathbf{x'} \geq 0,$$

with respect to the $l_2$ norm. In Eq. (6), $D_j \mathbf{x'}$ is the j$^{th}$ component of the discrete gradient of the vector $\mathbf{x'}$, and μ is the penalty parameter of the model (here we set μ=2$^6$). We note that for the reconstruction of the transmission function of the sample (the vector **x**) the vector $\mathbf{x'}$ that we obtained by using the described algorithm is renormalized to obtain

$$(7) \quad \mathbf{x}_j = \frac{x'_j}{\frac{1}{N}\sum_{k=1}^{N} A_{k,j}}.$$

The mathematical justification for this procedure is described in the supplemental information.

After we reconstructed separately the absorption spectrum for each SASE central energy of the SASE scan, we merged all the absorption spectra to create the spectrum of the sample. At this point, the number of data points is much larger than the number of points corresponding to the GS resolution since, as we will discuss below, the resolution of GS is determined by the width of the individual spectral spikes [41], which are broader than the resolution of the spectrometer. Therefore, the final step was to bin the points to obtain a bin size equal to the spike width.

### 3. Results and discussion

### 3.1 Ghost spectroscopy results

The GS results for the three samples are presented in figure 3. The blue dots are the GS reconstructions, and the magenta dots are the results of the synchrotron monochromatic scan that we used to validate our method. It is clear from figure 3 that the agreement between the GS reconstruction using the FEL and the synchrotron measurements is very good. We conclude that the energetic chemical shift of the Si $L_{2,3}$ resonances in the three different samples due to the different Si bonding is well monitored by GS reconstruction. Furthermore, the contribution of the spin orbit splitting to the $L_{2,3}$ edge is clearly visible in the absorption spectrum near 100.2 eV of the silicon membrane [42].

To compare the GS method, the scanning of the seeded radiation method, and the scanning of the SASE radiation method, we plot the spectra measured by these three different methods for the Si sample in figure 4(a).

While the photon energy resolutions of the seeded and the SASE radiation are simply determined by the spectral width of their pulse envelopes, the resolution of GS is mostly related to the spectral width of each individual spike of the SASE spectrum. More precisely, we measured the resolution of the three methods by calculating the FWHM of the autocorrelation function and by dividing it by $\sqrt{2}$. The auto-correlation function values of the SASE pulses averaged over 2000 pulses are the black dots in figure 4(b). The magenta line is a fit of the sum of two Gaussian curves. The two curves represent the spectral width of the SASE pulse and the width of the spectral spikes. Using the same procedure, we present the auto-correlation function of the seeded pulses in figure 4(c), where the narrower Gaussian corresponds to the spectral width of the seeded pulse. In figure 4(d) we present these three Gaussian fits that correspond to the energy resolutions of the SASE (green dashed line), seeded (red dotted line) and GS (blue solid line), respectively. It is important to note that the spectral width of the SASE radiation determines the spectral range of the GS reconstruction at each of the SASE central photon energies.

The results here indicate clearly that GS provides a higher resolution to the one reached in seeded mode and much higher than the SASE case, but with a number of scans that is equal to the number of scans used in SASE mode. For the same spectral range from 99 eV to 103 eV, with GS we used 15 steps and the resolution was 35 meV, with the SASE radiation we also used 15 steps, but the resolution was only 500 meV, and with the seeded radiation we used 51 steps and the resolution was 75 meV.

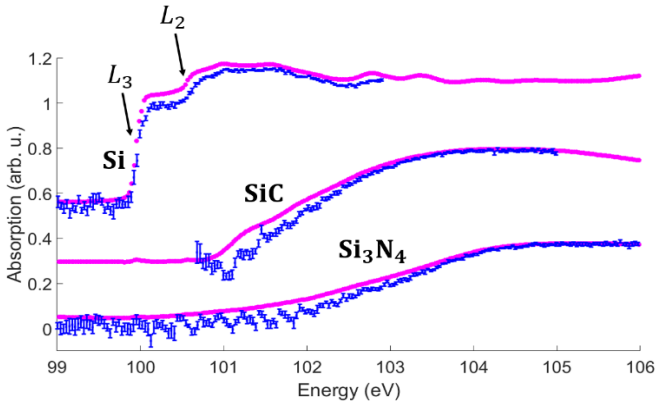

*Figure 3- GS $L_{2,3}$ edge spectra of silicon (Si), silicon carbonite (SiC) and silicon nitride ($Si_3N_4$). The blue dots are the GS results, and the magenta dots are the results obtained by the monochromatic synchrotron measurements.*

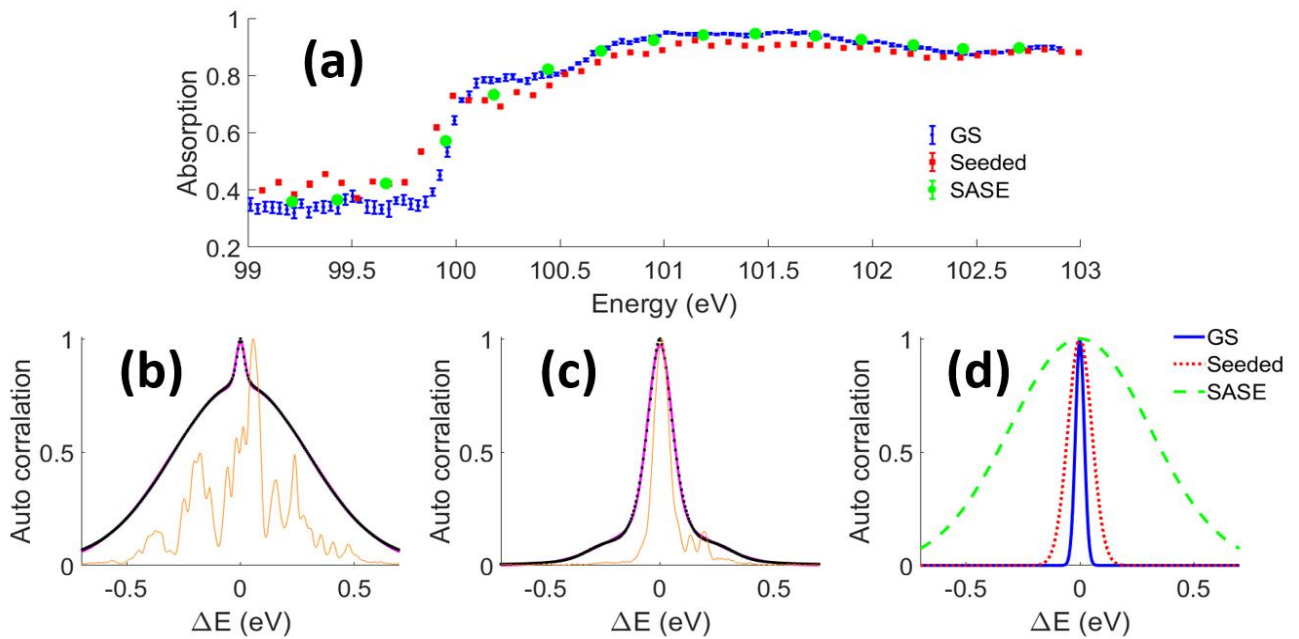

*Figure 4 – (a) Comparison of the GS results (blue dots) with the spectrum collected by scanning FEL pulses generated in the seeded (red squares) and the SASE (green circle) configurations. The statistical error of the SASE and seeded measurements was about 1% and too small to be seen. (b) and (c) Averaged auto-correlation function of SASE and seeded radiation, respectively. The black dots are the values of the auto-correlation function that are calculated from the measured data and the magenta line is a fit of sum of two Gaussians. The orange line is an example of a single shot spectrum. (d) Gaussian fits from the auto-correlation function, which we used for the calculation of the resolutions of the three methods. The resolutions are 500 meV, 75 meV, and 35 meV for SASE, seeded, and GS methods, respectively.*

In principle, it is possible to increase the spectral resolution in seeded mode by increasing the duration of seed pulses, but at the expense of increasing the number of scans to map a given spectral interval. For GS the trade-off between the resolution and the number of scans is lifted.

*3.2 Number of pulses required*

In the previous section, we showed that GS requires a smaller number of scans compared to scanning with the seeded radiation. However, the measurement time also depends on the number of shots required to stabilize the result at each energy point, which is different between the GS and the seeded beam scans.

A seeded FEL behaves indeed as laser like source [43] has a stable output wavelength and a stable output power. An ideal measurement of the input wavelength and of the pulse energy before and after the sample would be sufficient to measure one sample of the absorption spectrum with a single FEL shot. In practice, the noise associated to the energy detection reduces the correlation between the two energy measurements and even the acquisition of a single spectral sample requires averaging over a number of shots to reduce the measurement interval of confidence.

Conversely, in GS the spectrum reconstruction requires averaging over a set of spectral acquisitions. The single measurement provides only a fraction of the spectral information at each frequency sample, but simultaneously on the samples distributed over the broad range of frequencies corresponding to the SASE pulse bandwidth. The SASE spectrum has the structure of spikes of random amplitude and distribution, the fluctuation statistics of each bin in the measured spectrum depends therefore on the width of the bin itself. The highest achievable spectral resolution with GS corresponds to spectral bins separated by a width comparable to the spike spectral width. Reducing the spectral resolution by increasing the bin size reduces the spectral resolution, but improves the statistics collected at each shot and requires a lower number of shots to ensure the convergence of the analysis. It has been shown for GI that the number of iterations scales as the number of the pixels in the reconstructed image [44] and we expect a similar dependence for GS. To test this important aspect, we first compared the dependence of the quality of the absorption spectrum measurements on the number of shots for the GS and the seeded radiation.

To quantify the quality of the reconstructed spectrum we consider the synchrotron data as the accurate reference absorption spectrum and define the mean absolute error $\langle \varepsilon \rangle$ as:

(8) $$\langle \varepsilon \rangle = \frac{1}{M} \sum_{j=1}^{M} |x_j - g_j|,$$

where $x_j$ is the j$^{th}$ point in the reconstructed spectrum (by GS or seeded), $g_j$ is the j$^{th}$ point in the ground truth which is the synchrotron measurement, and M is the number of points.

We plot $\langle \varepsilon \rangle$ as a function of the number of shots for the GS (blue dots) and for the seeded radiation (red dots) in figure 5(a). The comparison (in addition to the synchrotron measurement as a reference) for various numbers of shots per scanning point are shown in figure 5(c-h). In our experiment the spectrum we reconstructed by GS was slightly closer to the synchrotron results than the spectrum obtained with the seeded radiation. The smallest $\langle\varepsilon\rangle$ for GS is around 0.035 and for the seeded radiation is 0.055. This comparison indicates that GS can provide high quality spectra with a number of pulses per point that is comparable to the seeded radiation scan.

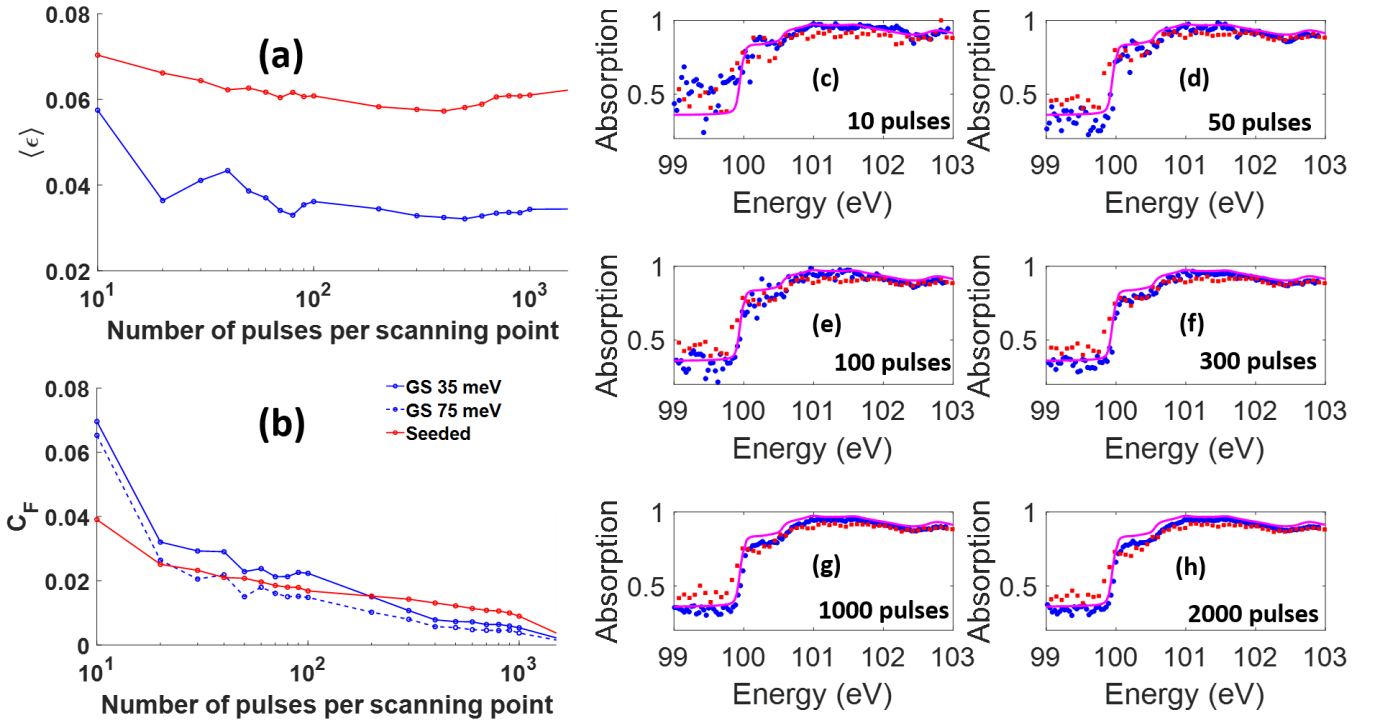

*Figure 5- (a) The mean absolute error and (b) Convergence factor of the reconstructed spectrum of the silicon membrane as a function of the number of pulses per scanning point for the GS (blue dots) and the seeded radiation (red dots). (c-h) The reconstructed spectrum results with GS (blue circles) and with seeded radiation (red squares) compared to the synchrotron measurements (magenta line) as a guide using (c) 10, (d) 50, (e) 100, (f) 300, (g) 1000, (h) 2000 pulses per scanning point.*

The small difference between the GS results and the seeded radiation results may indicate a systematic error and thus it is difficult to infer from figure 5(a) which method converges to its best value faster. We therefore defined the 'convergence factor', $C_F$, similarly to the above definition of the $\langle\varepsilon\rangle$ as:

(9) $\qquad C_F = \frac{1}{M}\sum_{j=1}^{M}|x_j - b_j|,$

where now $b_j$ represents the best value for GS and for the seeded mode separately, which is the j$^{th}$ point in the reconstructed spectrum using 2000 pulses for GS and seeded radiation. For this comparison we bin the GS dataset to 35 meV, corresponding to the value used in the plots (a)-(c-f) (blue-solid line) and 75 meV to match the resolution of the seeded radiation (blue-dashed line). In Fig 5(b) we show the $C_F$ for the GS and for the seeded radiation scans as a function of the number of pulses per scanning point in the three cases. The convergence of GS is comparable to the convergence of the seeded mode at equal resolution and slower when the resolution of the GS is higher, requiring more shots to reach a similar $C_F$. This implies that when the resolution is equal the required number of shots per data point with the GS and with the seeded radiation are comparable, hence the measurement time with GS is shorter than with the seeded radiation.

Another interesting result from this comparison is that even with a small number of 300 pulses per scanning point, the quality of the reconstructed spectrum obtained by GS was sufficient to resolve the main features of the Si L$_{2,3}$ edge spectrum. In our experiment the repetition rate was 50 Hz and the number of scanning points for the GS was 15, which implies that the measurement time was 90 seconds.

### 3.3 Compression factor dependence

As discussed in the previous sections the resolution of GS is determined by the average width of the single SASE spectral spikes. The main parameter that we can use to control the spike width is the electron bunch compression factor [45]. To compress the electron bunch, we used a magnetic double chicane located inside the acceleration section of the FERMI source. A typical compression is used at FERMI to increase the peak current up to about 650 A [46]. The compressor can be tuned to further compress the beam to reach a higher peak current. In figure 6(a) we show typical energy pulse spectra for a modestly compressed electron beam (blue line) and for a highly compressed electron bunch (purple dots), respectively. The two graphs clearly show that the primary spectral features are different in the two regimes, with more and wider spikes for a highly compressed electron beam.

To demonstrate this dependence more quantitatively we show in figure 6(b) the spike width (calculated by the FWHM of the auto-correlation Gaussians divided by $\sqrt{2}$) as a function of the peak current. The results indicate that the spectral spike width grows linearly with the compression of the electron beam, which is in agreement with the previous study [45]. In our experiment the peak current was about 900 A corresponding to a spectral width of 35 meV, while for higher current (i.e. 2150 A) the width is about 60 meV. The conclusion from this discussion is that by controlling the compression factor it is possible to control the spectral resolution of our approach. It is important to note however, that the compression also affects the coherence and the pulse duration of the SASE emission [45,47].

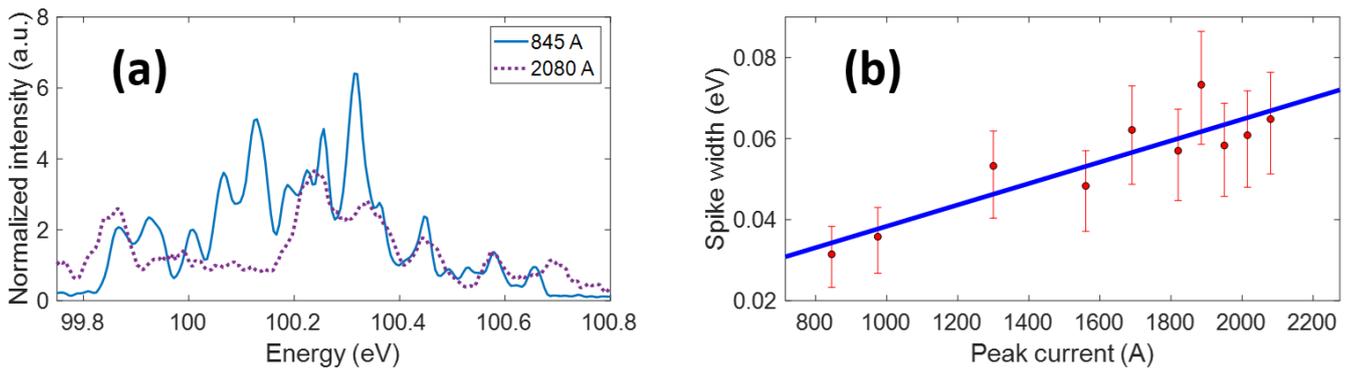

*Figure 6 - (a) Typical single shot FEL pulse spectra for 845 A (blue solid line) and 2080 A (purple dotted line) of the peak current of the electron bunch. (b) Spike width as a function of the peak current. The vertical error bars represent the uncertainty caused by the spectrometer resolution and the fitting procedure.*

## 5. Conclusion

In this paper we have demonstrated the implementation of GS for an XUV FEL with an on-line spectrometer in front of the sample and a photodiode after the sample. We have validated the quality of the absorption spectrum measurements by comparing to more conventional methods and found that the spectral resolution is comparable to the resolution of scans with seeded radiation. Our results indicate that the measurement time with GS can be significantly shorter than the measurement time with seeded radiation. From our analysis we conclude that the reduction in the measurement time is comparable to the ratio of the average spectral width of the SASE regime to the spectral width of the seeded one at a comparable spectral resolution. We emphasize that we do not expect that GS will be always faster than scans with quasi monochromatic radiation and stress that the exact conditions depend on the various details such as the specific properties of the FEL and the sparsity of the absorption spectrum. The spectral resolution of the GS method is determined by the lower resolution of either the width of spectral autocorrelation of the SASE radiation or the resolution of the spectrometer. Our work therefore calls for the improvement of the resolution of single-shot spectrometers and for the reduction of the spectral width of the SASE spikes. The extension of our work to transient spectroscopy measurements implemented by pump-probe approaches is straightforward and the reduction in the measurement time for those experiments will be more significant since the typical duration of pump-probe experiments is very long. However, we should emphasize that for pump-probe experiments also the temporal resolution should be considered and not just the spectral resolution. This is important since the temporal resolution is limited by the pulse duration that cannot be shorter than the Fourier transform limit of the single spike spectral width. Results of the pump-probe experiment investigated by GS and more in-depth discussion of the temporal effects will be given in a forthcoming paper.

## Acknowledgements

Y.K. gratefully acknowledges the support of the Ministry of Science & Technologies, Israel.

**Funding:** This work is supported by the Israel Science Foundation (ISF) (201/17);

Supplemental document for the paper: "High-resolution absorption measurements with free- electron lasers using ghost spectroscopy", Klein et al.

# RECONSTRUCTION PROCEDURE FOR GHOST SPECTROSCOPY WHEN USING SASE FLACTUATIONS

In the following paragraphs we provide further details on the derivation of the equations we used for the reconstruction of the absorption spectrum from the measured data in the experiment of ghost spectroscopy (GS).

To reconstruct the absorption spectrum of the object is we need to solve Eq. (2) of the main text:

(S1) $$\mathbf{T} = \mathbf{A}\mathbf{x},$$

where $\mathbf{T}$ is a vector that includes the intensities of N pulses measured by the photodiode, $\mathbf{x}$ is a P-length vector representing the transmission function of the sample, and $\mathbf{A}$ is a NxM matrix where every row represents the spectral distribution of a single pulse. However, the raw data from the measurements include not just the spectral response of the sample but also the constant envelope of the SASE fluctuations (the green line shown in figure 2(a) of the main text) and the shot-to-shot intensity instability, which is a general property of FELs. It is therefore necessary to eliminate the information that is not directly related to the sample. The idea of the procedure below is to exploit an auxiliary matrix $\mathbf{C}$ that contain the spectral information without the SASE envelope and to normalize the shot-to-shot intensity variation.

We recall that we can represent each element of the matrix $\mathbf{A}$ as:

(S2) $$A_{i,j} = a_i A^*_{i,j} F_j,$$

where the spectral envelope is the M-length vector F and the shot-to-shot intensity variation is represented by $a_i$ for each $i^{th}$ shot. $A^*_{i,j}$ is a random value from a normal distribution i.e. $A^*_{i,j} \sim N(\mu, \sigma^2)$. By inserting Eq. (S2) into Eq. (S1) we get:

(S3) $$T_i = \sum_{j=1}^{M} A_{i,j} x_j = \sum_{j=1}^{M} a_i A^*_{i,j} F_j x_j = \frac{a_i}{a} \sum_{j=1}^{M} A^*_{i,j} a F_j x_j,$$

where $a$ is defined as $a = \frac{1}{N}\sum_{i=1}^{N} a_i$. To use the matrix for the reconstruction of the absorption spectrum we define new matrix and vectors as:

(S4) $$x'_j = a F_j x_j \quad , \quad B_{i,j} = \frac{a_i}{a} A^*_{i,j} = \frac{A_{i,j}}{a F_j},$$

and Eq. (S3) can be transformed to:

(S5) $$T_i = \sum_{j=1}^{P} A'_{i,j} x'_j \quad \text{which is equivalent to} \quad \mathbf{T} = \mathbf{B}\mathbf{x}'.$$

The matrix B contains the data of the spectral distribution of the input pulses normalized by the average spectral envelope of the raw measurements.

Assuming $A^*_{i,j} \sim N(1, \sigma^2)$ and $a_i \sim N(a, \sigma^2)$, the envelope $aF$ can be calculated by averaging over all the raw pulses:

(S6) $$\frac{1}{N}\sum_{i=1}^{N} A_{i,j} = \frac{1}{N}\sum_{i=1}^{N} a_i A^*_{i,j} F_j = a F_j$$

Thus, the matrix B is related to the matrix A by:

(S7) $$\mathbf{B}_{i,j} = \frac{A_{i,j}}{aF_j} = \frac{A_{i,j}}{\frac{1}{N}\sum_{k=1}^{N} A_{k,j}},$$

which is Eq. (3) of the main text.

To address the second challenge of the shot-to-shot intensity variation we normalized each measurement $T_i$ by the value $a_i$ (and for convenience we multiply also by a constant factor $\frac{a}{M}$):

(S8) $$T'_i = \frac{a}{P}\frac{T_i}{a_i}$$

Using Eq. (S5) we can write

(S9) $$T'_i = \frac{a}{M}\frac{1}{a_i}\sum_{j=1}^{M} B_{i,j}\, x'_j.$$

and

(S10) $$C_{i,j} = \frac{a}{M}\frac{B_{i,j}}{a_i}.$$

This leads to

(S11) $$T'_i = \sum_{j=1}^{M} C_{i,j}\, x'_j \quad \text{or} \quad \mathbf{T'} = \mathbf{Cx'}.$$

which is Eq. (5) of the main text. The matrix **C** contains spectral data that we used for the reconstruction of the transmission function of the object. They are separated from the envelope of the measured spectra and normalized to have equal pulse energy. The factor $a_i$ is calculated by:

(S12) $$\frac{a}{M}\sum_{j=1}^{M} B_{i,j} = \frac{1}{P}\sum_{j=1}^{M}\frac{A_{i,j}}{F_j} = \frac{1}{M}\sum_{j=1}^{M}\frac{a_i A^*_{i,j} F_j}{F_j} = \frac{1}{M}\sum_{j=1}^{M} a_i A^*_{i,j} = a_i \frac{1}{M}\sum_{j=1}^{M} A^*_{i,j} = a_i.$$

The vector **T'** and the matrix **C** are related to the vector **T** and the matrix **B** by the relations

(S13) $$T'_i = \frac{a}{M}\frac{T_i}{a_i} = \frac{a}{M}\frac{T_i}{\frac{a}{M}\sum_{k=1}^{M} B_{i,k}} = \frac{T_i}{\sum_{k=1}^{M} B_{i,k}} \quad \text{and} \quad C_{i,j} = \frac{a}{M}\frac{B_{i,j}}{a_i} = \frac{B_{i,j}}{\sum_{k=1}^{M} B_{i,k}},$$

which lead to Eq. (4) of the main text. By using the definition of x' from Eq. (S4) we can find the actual transmission by using the relation

(S14) $$x_j = \frac{x'_j}{aF_j} = \frac{x'_j}{\frac{1}{N}\sum_{k=1}^{N} A_{k,j}},$$

which leads to Eq. (7) of the main text.